# Robust Wrapping-free Phase Retrieval Method Based on Weighted Least-square Method

Minmin Wang[a], Canlin Zhou*[a], Shuchun Si[a], XiaoLei Li[b], Zhenkun Lei[c], YanJie Li[d]

(a. School of Physics, Shandong University, Jinan , 250100, China

b.School of Mechanical Engineering, Hebei University of Technology, Tianjin, 300130, China

c. Department of engineering mechanics, Dalian University of Technology, Dalian,116024, China

d. School of civil engineering and architecture, Jinan University, Jinan, 2250020, China )

*Corresponding author: Tel: +8613256153609;    E-mail address: canlinzhou@sdu.edu.cn

## Abstract

For many profilometry techniques, phase unwrapping is one of the most challenging process. In order to sidestep the phase unwrapping process, Perciante et. al [Appl Opt 2015; 54(10):3018-23] proposed a wrapping-free method based on the direct integration of the spatial derivatives of the patterns to retrieve the phase. But it is only applicable for the case of the phase continuity for the tested object, which means it may fail to handle fringe patterns containing complicated singularities, such as noise, shadow, shears and surface discontinuity. In view of this problems, a robust wrapping-free phase retrieval method is proposed in this paper, which is based on combined Perciante's method and weighted least-squares method. Two partial derivatives of the desired phase is obtained from the fringe patterns, meanwhile the carrier is eliminated using direct phase difference method. The phase singularities are determined using derivative variance correlation map (DVCM), and the weighting coefficient is obtained from the binary mask of the reverse DVCM. Simulations and experiments are conducted to prove the validity of the proposed method. Results are analyzed and compared with those of Perciante's method, demonstrating that the proposed method can be available for measuring objects with some kinds of singularities sources.

## 1. Introduction

Phase-based fringe projection profilometry has been widely employed in three-dimensional shape acquisition due to its non-contact operation, simplicity, high resolution, reliability and fast data processing [1-4]. Over the years, many phase demodulation methods have been proposed to retrieve phase from fringe patterns. Some of these methods determine the phase by evaluating the arctangent of the ratio of the intensity functions, such as phase-shifting algorithm [5-7], Fourier-based techniques [8,9], and spatial interferometry [10]. So that the obtained phase is usually restricted in ($-\pi$, $\pi$]. The phase unwrapping must be carried out in these methods to estimate the continuous phase map from the wrapped ones. In practical applications, phase unwrapping may be quite difficult. For years, a great deal of techniques have been developed. As a general rule, most of these algorithms can be classified into two categories: path-following [11-13] and minimum-norm methods [14-17]. In the first class approaches, different quality maps, including correlation, pseudocorrelation, phase derivative variance, or maximum phase gradient, are possible means for helping with the path-selection problem. For instance, Aguemoune et. al [11] presented a phase unwrapping algorithm based on a contextual pseudocorrelation quality map and non-continuous path unwrapping. Wang et. al [12] used the mask produced from phase derivative variance as the cost function to unwrap the phase with discontinuities. Zhang et. al [13] proposed a phase unwrapping algorithm based on the quality map generated from the gradient of the phase map. For the other minimum-norm methods, minimum $L^p$-norm phase unwrapping algorithm, weighted and unweighted least-square phase unwrapping algorithm are typical examples exhibiting a good performance in the presence of noise. Ghiglia et. al [14] solved the phase unwrapping problem using the minimum $L^p$-norm solution, which is obtained by embedding the transform-based methods for least squares within a simple iterative structure. Ghiglia et. al [15] developed a least-square phase unwrapping algorithm for unwrapping wrapped phase by using fast cosine transforms. The weighted unwrapping problem is solved by iterations. Lu et. al [16] improved the weighted least-squares minimization methods by using the derivative variance correlation map (DVCM) as the weighting coefficient. It can truly reflect wrapped phase quality, ensuring a more reliable unwrapped result. Xia et. al [17] utilized iteration approach and calibration method to unwrap successfully the phase data corrupted by high speckle noise.

Apart from those methods mentioned above, many temporal phase unwrapping methods are proposed to unwrap the phase of the object with large discontinuities and isolated parts [18-24]. But they need many frames of fringe images which would take much time. The phase retrieval method based on the polynomial fitting technique [25-26] is also presented to estimate directly the continuous phase. But how to find the appropriate polynomial base is not a trivial problem. Pandey et. al [27] proposed the phase unwrapping method based on the transport of intensity equation (TIE). The phase is automatically in the continuous form as it is a solution of a partial differential equation rather than as an argument of arctangent function. However, Pandey's method will not work satisfactorily when the phase map has a curl component.

Many scholars presented the other phase retrieval methods which completely sidestep the phase unwrapping process. For instance, Zhong [28] introduced a fast phase measurement profilometry based on phase-shifting and trifocal tensor. Lohry [29] modified fringe patterns by encoding the quality map for efficient and accurate stereo matching. Tao et. al [30] introduced a high-speed 3-D shape measurement technique based on composite phase-shifting fringes and a multi-view system. Gai et. al [31] proposed an efficient three-dimensional shape measurement system based on the combining projection of single digital speckle pattern and phase-shifting fringe patterns. Neither of them are necessary to obtain the unwrapped phase. However, they are all based on the principle of binocular stereoscopic vision and feature high hardware complexity. References [32,33] described a method of determining phase through directly obtaining and integrating the derivatives of the desired phase, avoiding a phase unwrapping process, but it needs to choose carefully the integral path for the case that there are discontinuities (e.g. holes) in fringe patterns.

Recently, Perciante et. al [34] presented a wrapping-free phase retrieval method. It determines the phase of the test object following two steps: first, it derives two partial derivatives of the desired phase from the fringe patterns, and then it reconstructs the phase by the line integral of the two partial derivatives. Different from the method by reference [28-33], it transforms the direct integration of the spatial derivatives to solve Poission's equation. However, Perciante's method has a single prerequisite that the phase of the tested object is continuous. In other word, the method is not capable of measuring objects containing some singularities, such as noise, shears, shadows and inconsistency (e.g. holes, steep surface) .

To solve the problems in Perciante's method, here, we present a robust wrapping-free phase retrieval method, which is based on combined Perciante's method and weighted least-squares method. Two partial derivatives of the desired phase is obtained from the fringe patterns, meanwhile the carrier is eliminated using direct phase difference method. The phase singularities are determined through the DVCM, which means the weighting coefficient is obtained from the binary mask of the reverse DVCM. Because the weighting coefficients corresponding to the singularities are given zero-weighted, the singularities won't affect the unwrapping. We tested the method we are proposing with simulations and experiments, and compared it with Perciante's method. The results demonstrate that the proposed method can successfully be used to measure objects containing some kinds of singularities. They illustrate the validity and superiority of the proposed method.

The rest of the paper is organized as follows. Section 2 describes the theory. Section 3 presents the computer simulation results. Section 4 shows experimental validation. Section 5 summarizes the full paper.

## 2. Theory

*2.1 Perciante's method*

The wrapping-free phase retrieval method proposed by Perciante et. al has been thoroughly described and discussed [34], this section only introduces its main idea briefly. Suppose Perciante's method is applied to four-step phase-shifting fringe. The deformed fringe patterns are mathematically described as:

$$I_k(x,y) = A(x,y) + B(x,y)\cos[2\pi fx + \varphi(x,y) + k\pi/2] \quad (k = 0, 1, 2, 3) \quad (1)$$

where $(x, y)$ is Cartesian coordinates, $A(x, y)$ the background intensity, $B(x, y)$ the intensity modulation amplitude, $\varphi(x, y)$ the shape-related phase and $2\pi fx$ the carrier fringe-related phase components. As for those conventional methods, the wrapped phase is retrieved by an expression below.

$$\varphi_w(x,y) = \tan^{-1}\left[\frac{S(x,y)}{C(x,y)}\right] \quad (2)$$

where $S(x, y) = I_3(x, y) - I_1(x, y)$ and $C(x, y) = I_2(x, y) - I_4(x, y)$. In this way, the phase unwrapping must be carried out to obtain actual continuous phase. However, to avoid such phase unwrapping process, Perciante's method determines the phase by two steps: firstly, it derives two partial derivatives of the desired phase from the fringe patterns; secondly, it solves Poission's equation to estimate the continuous phase.

The partial derivatives of the phase with respect to spatial coordinates are

$$g \equiv \frac{\partial \varphi_w}{\partial x} = \frac{C(\partial S / \partial x) - S(\partial C / \partial x)}{S^2 + C^2} \quad (3)$$

$$h \equiv \frac{\partial \varphi_w}{\partial y} = \frac{C(\partial S / \partial y) - S(\partial C / \partial y)}{S^2 + C^2} \quad (4)$$

Then the unwrapped phase $\phi_c(x,y)$ can be obtained by getting the least-square solution, which can be obtained as the solution of the following Poisson equation:

$$\frac{\partial^2 \varphi_c}{\partial x^2} + \frac{\partial^2 \varphi_c}{\partial y^2} = \frac{\partial g}{\partial x} + \frac{\partial h}{\partial y} \quad (5)$$

Finally, the shape-related phase $\varphi(x,y)$ is obtained after the phase of the reference plane is subtracted.

*2.2 Our method*

Perciante's method can retrieve the phase under the single assumption that the phase is continuous and reliable, which means it may cause failure when dealing with phase with singularities. In an attempt to solve this problem, here, we present a robust wrapping-free phase retrieval method, which is based on combined Perciante's method and weighted least-squares method

Firstly, the direct phase difference method [35] is introduced in to remove the carrier conveniently. Apart from the set of fringe patterns deformed by the measured objects (Eq. (1)), we capture the other set of patterns by reference plane, as shown in Eq. (6).

$$BI_k(x,y) = A(x,y) + B(x,y)\cos[2\pi f x + k\pi/2] \quad (k = 1, 2, 3, 4) \quad (6)$$

The phase difference is calculated directly using two sets of patterns, which means the functions $S(x, y)$ and $C(x, y)$ are redefined as shown in Eqs. (7) and (8).

$$S_p = (I_4 - I_2) \times (BI_1 - BI_3) - (BI_4 - BI_2) \times (I_1 - I_3) \quad (7)$$

$$C_p = (BI_4 - BI_2) \times (I_4 - I_2) - (BI_1 - BI_3) \times (I_1 - I_3) \quad (8)$$

$S_p(x, y)$ and $C_p(x, y)$ are then utilized to obtain the two partial derivatives of the shape-related phase $\varphi(x, y)$ in similar way as Eqs. (3) and (4).

$$g \equiv \frac{\partial \varphi}{\partial x} = \frac{C(\partial S_p / \partial x) - S(\partial C_p / \partial x)}{S_p^2 + C_p^2} \quad (9)$$

$$h \equiv \frac{\partial \varphi}{\partial y} = \frac{C(\partial S_p / \partial y) - S(\partial C_p / \partial y)}{S_p^2 + C_p^2} \quad (10)$$

After obtain the two partial derivatives of the phase, we introduce the weighted least-square phase retrieval method [15,16] to retrieve the continuous phase effectively. In weighted least-square algorithm, how to design the weighting coefficient is important. Generally, the quality map is used to define the weighting coefficient. Among all the quality maps, the DVCM is widely recognized as the most reliable map, and we will use it to detect and mask the phase singularities here. But we find that the DVCM defined in Ref. 16 represents the "badness" rather than the "goodness" of the phase, as its values in the areas with "badness" of the data are bigger than those in the areas with goodness of the data, which means its influence of the "badness" is magnified, while the good reliable values are neglected. So that it will fail when directly using the square of DVCM as the weighed coefficient to retrieve the phase.

Here, we propose to take another quality map, namely reverse DVCM as shown in Eq. (11), which is directly derived from the fringe patterns intensity.

$$q(x,y) = 1 - \frac{\sqrt{\sum_{i=x-l/2}^{x+l/2}\sum_{j=y-l/2}^{y+l/2}[g(i,j)-\bar{g}(x,y)]^2} + \sqrt{\sum_{i=x-l/2}^{x+l/2}\sum_{j=y-l/2}^{y+l/2}[h(i,j)-\bar{h}(x,y)]^2}}{l \times l}$$
$$\times \left(1 - \frac{\sqrt{\left[\sum_{i=x-l/2}^{x+l/2}\sum_{j=y-l/2}^{y+l/2}c(i,j)\right]^2 + \left[\sum_{i=x-l/2}^{x+l/2}\sum_{j=y-l/2}^{y+l/2}s(i,j)\right]^2}}{l \times l}\right) \quad (11)$$

where $c = -C_p/\sqrt{C_p^2 + S_p^2}$ and $s = -S_p/\sqrt{C_p^2 + S_p^2}$. For each sum the indexes $(i, j)$ range over the $l \times l$ window centered at the pixel $(x, y)$. The terms $g(i,j)$ and $h(i,j)$ are the partial derivateives of the shape-related phase $\varphi(x, y)$, described as Eqs. (9) and (10).. The terms $\bar{g}(x,y)$ and $\bar{h}(x,y)$ are averages of $g(i,j)$, $h(i,j)$ in the $l \times l$ windows.

However, directly using the square of reverse DVCM as the weighting coefficient, we find out that the weighted least squares algorithm cannot retrieve a very reliable continuous phase after trial and error. Here, we generate a mask whose pixels take only two values: 0 and 1 through the binary of the square of reverse DVCM. To calculate the mask, a threshold must be selected appropriately. Here, we employ Otsu's method [36], which requires no manual intervetnion. Any values bellow the threshold are set to 0, while those above the threshold are set to 1. Suppose the obtained threshold is $t_{otsu}$, the binary of the square of reverse DVCM can be calculated as

$$M(x,y) = \begin{cases} 1 & q^2(x,y) > t_{otsu} \\ 0 & q^2(x,y) \le t_{otsu} \end{cases} \quad (12)$$

The mask is used to define the weighting coefficient. The zero-valued pixels represent the low-quality values that are to be zero-weighted and ignored. Therefore, our method is robust in the presence of many kinds of phase singularities, and able to handle these cases as zero-weights are employed to mask out the region with singularities.

After taking into account the weighting coefficient $w(x,y) = M(x,y)$, the Poisson equation of Eq. (5) is re-expressed as

$$\begin{aligned} &w(x,y)[\phi(x+1,y)-\phi(x,y)] + w(x-1,y)[\phi(x-1,y)-\phi(x,y)] \\ &+w(x,y)[\phi(x,y+1)-\phi(x,y)] + w(x,y-1)[\phi(x,y-1)-\phi(x,y)] \\ &= w(x,y)g(x,y) - w(x-1,y)g(x-1,y) + w(x,y)h(x,y) - w(x,y-1)h(x,y-1) \end{aligned} \quad (13)$$

where $\phi(x,y)$ is the unwrapped shape-related phase. The preconditioned conjugate gradient

(PCG) technique [16,37,38] is applied to solve the weighted least squares described by Eq. (13). Finally, the wrap-free phase, which is immune to the influence of some kinds of singularities is recovered.

## 3. Numerical simulation

In order to testify the feasibility and advantages of the proposed method, we carried out numerical simulations. The phase distribution of test surface generated by Matlab is given by Eq. (14), and its 3D surface is show in Fig. 1(a), at a resolution of 512 × 512 pixels.

$$\varphi = [(x-256)/50]^2 + [(y-256)/50]^2 \tag{14}$$

Two sets of fringe patterns were produced according to Eqs. (1) and (6). In order to figure out the influence of singularities, such as shadows on both Perciante's method and the proposed method, the rectangular regions in these fringe patterns were padded with values close to 0. One of the fringe patterns deformed by the test sample ($I_1$) is shown in Fig. 1(b).

We firstly utilized Perciante's method to process the simulated patterns depicted in Fig. 1(b). The recovered phase is shown in Fig. 1(c), from which the carrier is particularly removed by subtracting the reference plane. The root mean square errors (RMSE) between the theoretical value and the retrieved phase is 1.0843 rad. It is easy to see that Perciante's method accommodates noise and the unwrapped phase in the rectangular region is greatly destroyed. Then we adopted the proposed method. $S_p(x, y)$ and $C_p(x, y)$ were determined using Eqs. (7) and (8) to remove the carrier, and two partial derivatives of the phase were obtained from Eqs. (9) and (10). If we use the square of reverse DVCM as the weighting coefficient to solve Eq. (13) using weighted least-square phase unwrapping algorithm, the recovered wrapping-free phase is still damaged by large noise with the RMSE of 0.4209 rad, as shown in Fig. 1(d). Instead, calculating the binary mask from Eq. (12) as the weighting coefficient and solving Eq. (13), we obtain the phase shown in Fig. 1(e). In this way, the RMSE is 0.2016 rad. Clearly, the error of the restored phase related to the shadow areas is reduced.

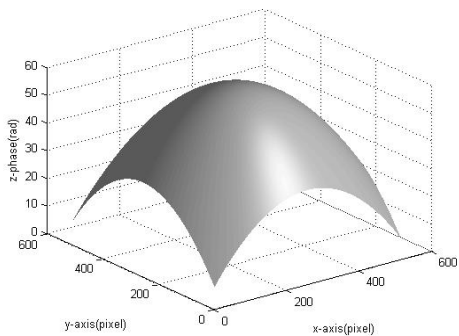
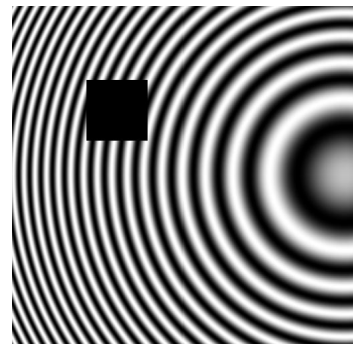

(a)　　　　　　　　　　　　　　(b)

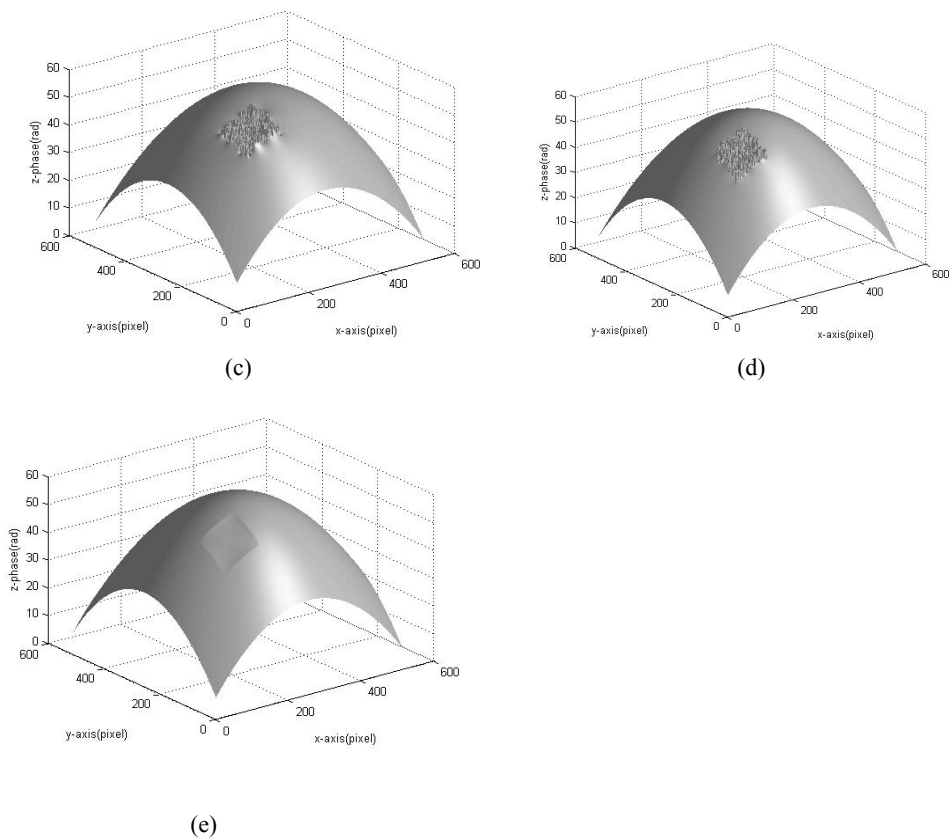

(e)

Fig. 1 Simulation results about shadows; (a) Phase distribution of tested sample; (b) One of the deformed fringe patterns; (c) Unwrapped phase obtained using the Perciante's method; (d) Unwrapped phase obtained using square of reverse DVCM as the weighting coefficient; (e) Unwrapped phase obtained using binary mask as the weighting coefficient.

Another two sets of fringe patterns discontinued with a shear along several horizontal lines between the top and the bottom of the image were generated to further compare Perciante's method and the proposed method. One of the patterns deformed with test surface is shown in Fig. 2(a). The process is similar to the previous simulation. The retrieved phase using Perciante's method and proposed method (regard the binary mask as the weighed coefficient) is respectively shown in Figs. 2(b) and 2(c), the RMSE of which is 0.4537 rad and 0.1747 rad, respectively. The results demonstrate that the proposed method is more robust to such discontinuity than Perciante's method.

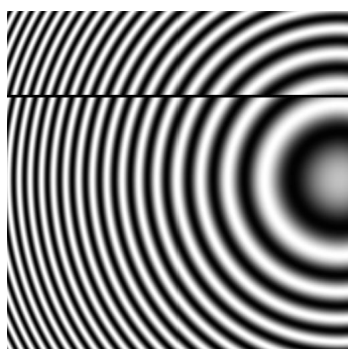  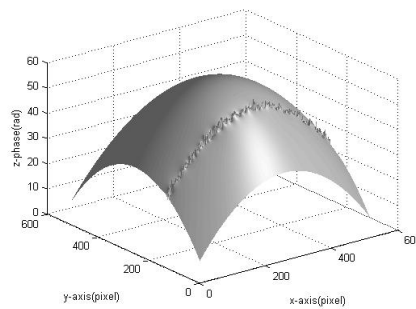

(a)  (b)

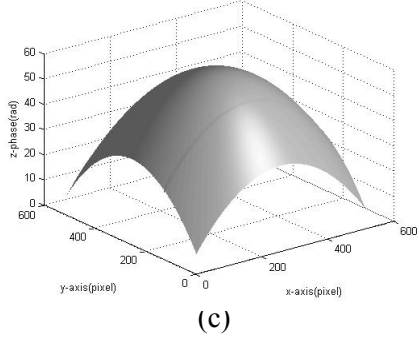

(c)

Fig. 2 Simulation results about discontinuity; (a) One of the deformed fringe patterns; (b) Unwrapped phase obtained using Perciante's method; (c) Unwrapped phase obtained using the proposed method.

## 4. Experiment and result analysis

Next, we developed a fringe projection measurement system to further verify the performance of the proposed algorithm, which consists of a DLP projector (Optoma DN344) and a CCD camera (DH-SV401FM). The camera uses a 25 mm focal length mega-pixel lens (ComputarFAM2514-MP2) with a 768 × 576 pixels resolution, and a maximum frame rate of 50 frames/s. The measurement software is programmed in MATLAB.

The tested object is a face module. Two sets of patterns ($I_1, \ldots I_4$ and $BI_1, \ldots BI_4$) when four phase-shifted sinusoidal fringe patterns were projected on a module and reference plane were captured by the camera, as shown in Figs. 3(a) and 3(b). The quality of those values around partial edges of the face is lower than others. The retrieved phase using Perciante's method (after subtraction of the reference plane) is shown in Fig. 3(c). Red circles indicate where the noisy regions of data lines.

Then we employed the proposed method. The functions $S_p(x, y)$, $C_p(x, y)$ and two partial derivatives of the phase were calculated directly using fringe patterns to remove the carrier frequency. Then the weighting coefficient is calculated. Fig. 3(d) shows the square of reverse DVCM. Utilizing Fig. 3(d) as the weighting coefficient, though the quality of recovered phase (Fig. 3(e)) has slightly improved in the discolorations areas, a large amount of noise is still there on the side of the head. Then we solved the binary mask, as shown in Fig. 3(f). Applying Fig. 3(f) as the weighting coefficient and utilizing the weighted least-square phase unwrapping method, we obtained the wrapping-free phase as shown in Fig. 3(g). Clearly, the quality and continuity of the retrieved phase map is much better than Figs. 3(c) and 3(e), especially in the corresponding discolorations areas.

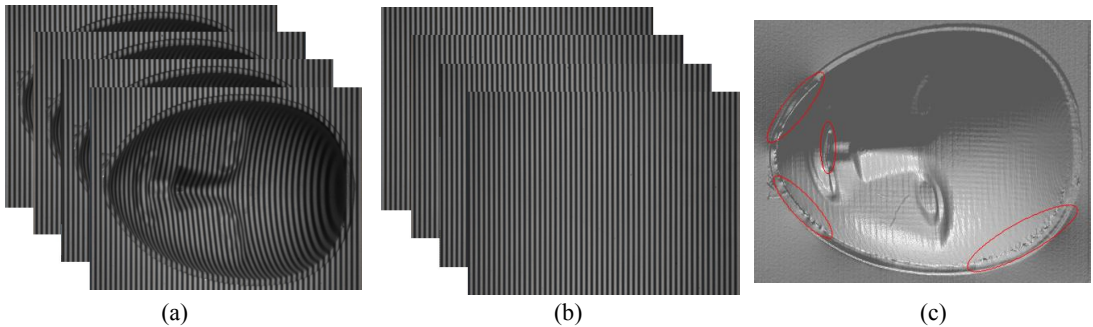

(a) (b) (c)

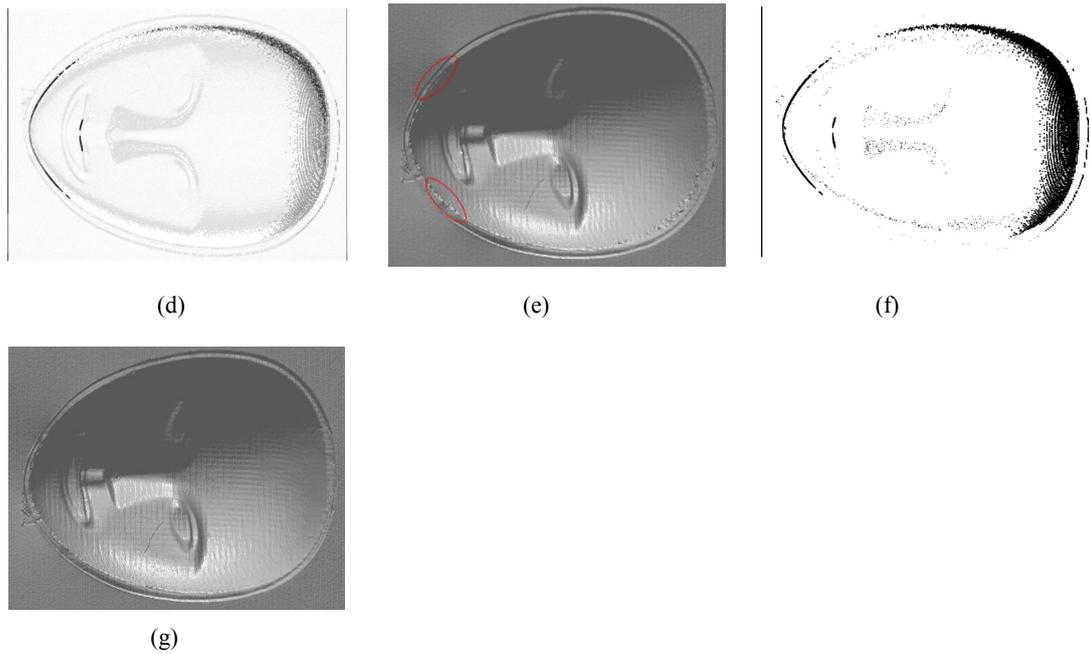

(d)            (e)            (f)

(g)

Fig. 3 Experimental results: a face module; (a) - (b) Four patterns acquired when four π/2-shifted sinusoidal fringe patterns were respectively projected on a module and reference plane; (c) Phase map obtained using Perciante's method; (d) Square of DVCM calculated using Eq. (11); (e) Phase map obtained using Fig. 3(d) as the weighting coefficient; (f) Binary mask calculated using Eq. (12); (g) Phase map obtained using Fig. 3(f) as the weighting coefficient.

Subsequently, two isolated objects, the face module and half of a paper cup, which has obvious change of height was measured. There is obvious shadow at the top of the cup. Besides, to test the robustness of the method, invalid data covered both sides of the fringe patterns, as shown in Figs. 4(a) and 4(b). Fig. 4(c) shows the retrieved phase of measured objects using Perciante's method. The phase have obvious undulation and noise in the discolorations areas. Fig. 4(d) shows the retrieved phase utilizing the square of reverse DVCM as the weighed coefficient. The phase still has large noise though it is a little bit better than Fig. 4(c). Fig. 4(e) shows the phase recovered using the proposed method, in which the binary mask is employed as the weighting coefficient. Obviously, the proposed method achieves comparative accuracy to the previous method.

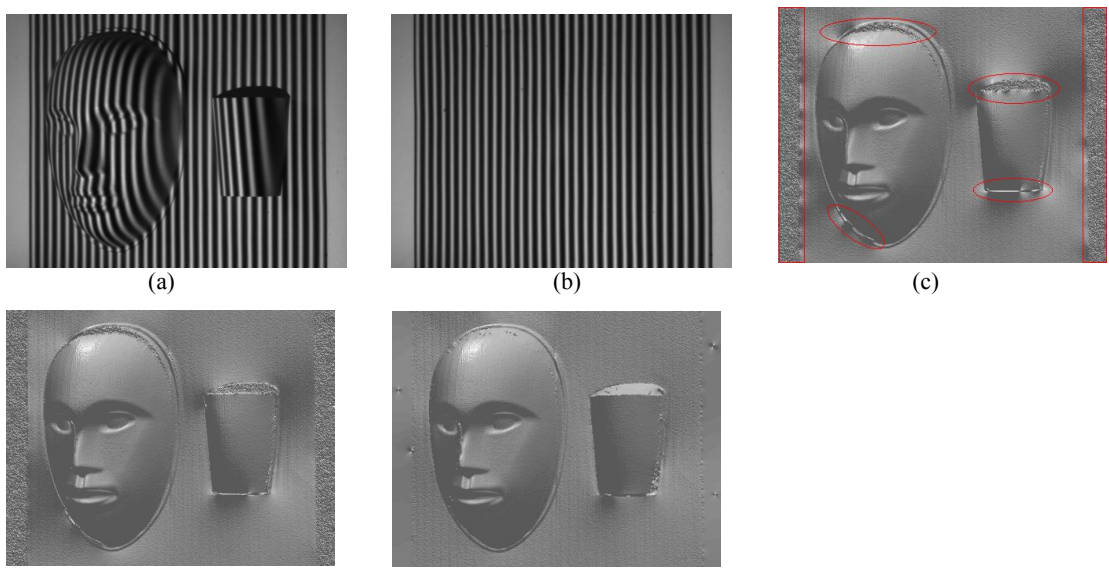

(a)            (b)            (c)

(d)                    (e)

Fig. 4 Experimental results: a face module and half of a cup; (a) One of the patterns acquired when sinusoidal fringe patterns were projected on a module; (b) One of the patterns acquired when sinusoidal fringe patterns were projected on the reference plane; (c) Phase map obtained using Perciante's method; (e) Phase map obtained using square of reverse DVCM as the weighting coefficient; (g) Phase map obtained using binary mask as the weighting coefficient.

## 5. Conclusion

In this paper, we present a robust wrapping-free phase retrieval method based on weighted least-square method. It is an extension of Perciante's method. Compared with Perciante's method, the proposed method not only retains the advantage of phase determination by direct integration of the spatial derivatives of the patterns, but also breaks the limitation that it is only applicable for measuring continuous surface. By introducing the weighed least-square method and derivative variance correlation map, a weighting coefficient is determined from the binary mask through the reverse DVCM and Otsu's automatic thresholding. Therefore the proposed method can detect and mask out many kinds of phase singularities, and it is immune to the destructive influence of some kinds of singularities. Simulations and experiments demonstrate the suitability and superiority of the approach.

**Acknowledgment**
This work was supported by the National Natural Science Foundation of China (Grant nos: 11672162,11302082 and 11472070). The support is gratefully acknowledged.